%% file: main.tex
\documentclass[acmlarge]{acmart}

\usepackage{natbib}
\usepackage{makecell,xspace}
\newcommand*{\ie}{i.e.,\xspace}
\newcommand*{\eg}{e.g.,\xspace}

\newcommand{\rev}[1]{\textcolor{black}{#1}}
\newcommand{\eleni}[1]{\textcolor{black}{#1}}

\begin{document}

%%
%% The "title" command has an optional parameter,
%% allowing the author to define a "short title" to be used in page headers.
\title{Towards Substantive Conceptions of Algorithmic Fairness: Normative Guidance from Equal Opportunity Doctrines}

%%
%% The "author" command and its associated commands are used to define
%% the authors and their affiliations.
%% Of note is the shared affiliation of the first two authors, and the
%% "authornote" and "authornotemark" commands
%% used to denote shared contribution to the research.

%\author{Anonymous Authors}

\author{Falaah Arif Khan}
\affiliation{%
  \institution{New York University}
  \country{New York, USA}}
 \email{fa2161@nyu.edu}

 \author{Eleni Manis}
 \affiliation{%
  \institution{Surveillance Technology Oversight Project }
  \country{New York, USA}
}

\author{Julia Stoyanovich}
\affiliation{%
  \institution{New York University}
  \country{New York, USA}
}

%%
%% By default, the full list of authors will be used in the page
%% headers. Often, this list is too long, and will overlap
%% other information printed in the page headers. This command allows
%% the author to define a more concise list
%% of authors' names for this purpose.
%\renewcommand{\shortauthors}{Arif Khan, et al.}

%%
%% The abstract is a short summary of the work to be presented in the
%% article.
\begin{abstract}
In this work we use Equal Oppportunity (EO) doctrines from political philosophy to make explicit the normative judgements embedded in different conceptions of algorithmic fairness. We contrast formal EO approaches that narrowly focus on \emph{fair contests} at discrete decision points, with substantive EO doctrines that look at people's \emph{fair life chances} more holistically over the course of a lifetime. We use this taxonomy to provide a moral interpretation of the impossibility results as the incompatibility between different conceptions of a \emph{fair contest} --- foward-facing versus backward-facing --- when people do not have \emph{fair life chances}. We use this result to motivate substantive conceptions of algorithmic fairness and outline two plausible \emph{fair decision procedures} based on the luck-egalitarian doctrine of EO, and Rawls's principle of fair equality of opportunity. 
\end{abstract}

%%
%% This command processes the author and affiliation and title
%% information and builds the first part of the formatted document.
\maketitle

\input{eo}
\input{notation}

\input{formal}

\input{formal_plus}

\input{impossibility}

\input{substantive}
\input{taxonomy}
\input{normative}

\input{related_work}
\input{conclusion}

\bibliography{main}

\end{document}

%% file: eo.tex
\section{Equality of Opportunity}
\label{sec:eo}

Equality of Opportunity (EO) is a philosophical doctrine that objects to morally arbitrary and irrelevant factors affecting people's access to desirable positions, and the social goods attached to them (such as opportunity and wealth). In an EO-respecting society, all people, irrespective of their morally arbitrary characteristics, such as socio-economic background, gender, race, or disability status, have comparable access to the opportunities that they desire. Similarly, in fair machine learning (fair-ML), we are usually interested in ensuring that the outputs of algorithmic systems, specially those used in critical social contexts, do not systematically skew along the lines of membership in protected groups based on gender, race, or disability. In so far as protected groups are constructed on the basis of morally arbitrary factors, the moral desiderata of EO doctrines from political philosophy align exactly with the fairness-related concerns in machine learning. In this work, we employ ideas from the rich literature on Equality of Opportunity from political philosophy \cite{anderson_integration, Anderson1999Equality, Arneson2018FourCO, Fishkin2014Bottlenecks, dworkin_1981,Lefranc_Trannoy, Rawls1971Justice, Roemer2002, williams_1973, young1994equity, jencks} to clarify the normative foundations of fairness and justice-related interventions, and gauge the efficacy of current algorithmic approaches that attempt to codify these criteria. 

\subsection{Principles of EO}
\label{sec:eo:principles}

There are two broad principles of EO, namely, \emph{the principle of fair contests} and \emph{the principle of fair life chances}.

\subsubsection{Fair contests}
The principle of fair contests, commonly understood as the \emph{nondiscrimination principle}, says that competitions for desirable positions should \eleni{ be open to all and should be adjudicated based on competitors' relevant merits, or qualifications.} In any fair contest, the most qualified person wins.  Conversely, fair contests do not judge \eleni{competitors on the basis of irrelevant characteristics, especially excluding} morally arbitrary factors such as gender, race, and socio-economic status, \eleni{that are not properly understood as qualifications at all}.

The principle of fair contests has been very influential in fair-ML and has guided statistical measures and algorithmic interventions that conceptualize \emph{fairness} as \emph{nondiscrimination}.

\subsubsection{Fair life chances}
The principle of fair life chances says that people's chances of success over a lifetime should not depend on morally arbitrary factors. The principle of fair life chances takes a holistic view of equal opportunity by comparing the \emph{opportunity sets} that people have over the course of a lifetime, and is popularly understood as a principle that \emph{levels the playing field}.

The principle of fair life chances has been almost entirely overlooked in fair-ML, and this omission explains some of the limitations in current approaches, as we will discuss shortly.

\subsection{Domains of EO}
\label{sec:eo:domains}
According to \citet{Fishkin2014Bottlenecks}, there are broadly three domains of equal opportunity:

\subsubsection{Fairness at a specific decision point} The first domain comprises of the discrete points at which social goods are distributed, such as employment, admissions and loan decisions. EO doctrines compel us to think about whether outcomes of decision-making at discrete decision points are influenced by morally arbitrary factors. 

\subsubsection{Equality of developmental opportunities} The second domain comprises educational and other foundational opportunities that shape people's ability to compete for desirable positions in the first domain. EO doctrines are also concerned with whether people had comparable developmental opportunities to build up their qualifications ahead of competitions.

\subsubsection{Equality of opportunities over the course of a lifetime} Lastly, EO doctrines also compel us to look more broadly at the \emph{opportunity sets} that people have access to over the course of a lifetime, and whether these bundles are comparable.

\subsection{Roadmap}
\rev{Different interpretations of the two principles discussed in Section~\ref{sec:eo:principles}, and to which domain they apply, give rise to different conceptions of EO. We fix notation in Section~\ref{sec:notation}.  We then discuss Formal doctrines --- which emphasize the principle of \emph{fair contests} at a discrete decision points --- in Sections~\ref{sec:formal} and~\ref{sec:formal_plus}. We then provide a moral interpretation of the impossibility results in fair-ML as the incompatibility between a forward vs. backward-facing conception of a fair contest in Section \ref{sec:impossibility}. In Section \ref{sec:substantive} we discuss Substantive EO doctrines --- which emphasize the principle of \emph{fair life chances} and target all three domains --- as they are classically understood. Next, in Section \ref{sec:substantive-modern}, we provide modern re-interpretations of these doctrines that are more amenable to real-world decision-making. In Section~\ref{sec:taxonomy}, we arrange the moral desiderata of different EO doctrines into a \emph{Fairness as Equal Opportunity} taxonomy. In Section \ref{sec:normative} we demonstrate how our EO-based framework can provide normative guidance in practical contexts using a hypothetical example of college admissions, and the real-world case study of COMPAS. We compare our framework with contemporary work in Section \ref{sec:related_work}, and then conclude with a discussion about the importance of grounding algorithmic approaches in strong normative foundations, as well as the limitations in guidance that EO doctrines can provide us towards this end.}

%% file: notation.tex
\section{Notation}
\label{sec:notation}

Before discussing different EO doctrines and their codification in fair-ML, let us fix some notation: Algorithmic decision-making involves predicting an outcome $y'$ given a set of observations $(X, y)$, where the $X$s are covariates/features and the $y$s are the targets. In a given context, we assume that covariates $X$ can be partitioned into $A$ ``morally relevant'' attributes, and $S$ ``morally arbitrary'' attributes (such as gender, race, age, disability, etc). ``Fair'' decision-making is concerned with satisfying some moral desiderata $C$, with respect to the set of attributes $S$, to which we commonly refer as the ``sensitive attributes''. 

We can apply EO doctrines to predictive problems if the target $y$ indicates a social outcome; where people receive or don't receive some desirable social good, the covariates $A$ measure some notion of merit, and protected-groups are constructed on the basis of a morally arbitrary and irrelevant characteristic \eleni{or characteristics} $s \in S$. For example, predicting the risk of loan default can be posed as the problem of allocating a positive or negative lending decision, and the co-variates measure ``financial qualifications'' like income, pay behavior, etc. Racial discrimination is illegal in lending, and so the morally arbitrary feature set in this example would include race. 

For the rest of the discussion, we restrict ourselves to decision-making of this type, where the predicted outcome $y'$ is a real valued score used to make a distributive/allocative decision. \emph{We stress here that EO doctrines are only suitable to predictive contexts that are amenable to such a translation, \ie they can be posed as the problem of distribution of some social good, on the basis of some relevant qualifications/merit.} We emphasize the non-applicability of EO doctrines to algorithmic contexts that cannot be posed as distributive problems \cite{Rawls1971Justice}.

%% file: formal.tex
\section{Formal EO}
\label{sec:formal}

Formal EO doctrines specify the moral desiderata of \emph{fair contests}: they say that no person should be excluded from a competition for a desirable position on the basis of morally arbitrary criteria. %, such as gender, race, disability, etc.
Further, in a \emph{fair contest}, people should only be judged on the basis of their relevant merit, and so, people with comparable relevant qualifications should get the same outcome. 

Formal EO, commonly known as \emph{careers open to talents}, is only concerned with the first domain of equal opportunity, \ie fairness at a discrete decision point. Formal EO is not attentive to whether people had comparable access to developmental opportunities to build qualifications leading up to the \emph{fair contest} (the second domain), nor whether people will have comparable opportunity sets over the course of their lifetimes (the third domain). In Bernard Williams's famous example of a warrior society, formal EO is achieved when \eleni{warrior} positions are open to all, and all are allowed to compete --- not just the children of warrior parents ~\cite{williams_1973}. However, formal EO fails to prevent privilege from being converted into qualifications in advance of the competition, whereby children \eleni{from non-warrior families} have no realistic chance of winning the contest without the resources and training that is afforded to children of warrior parents.

\subsection{Formal EO as Fairness through Blindness}
Decision-making that is blind to irrelevant characteristics is consistent with formal EO. In fair-ML, a prominent codification of formal EO is \emph{fairness through blindness}~\cite{dwork_awareness}, where protected (and morally irrelevant) attributes %such as gender, race, disability status, etc 
are removed from the data, and a group-blind classifier is produced.  To the extent that irrelevant characteristics (and their proxies) can be successfully excluded from an algorithm’s pipeline, formal EO can make progress toward its aim of rejecting the use of morally irrelevant features as the basis for awarding privileged outcomes. In practice, however, formal EO is often too weak to have a fairness-enhancing effect, as has been demonstrated in both the digital and analog age~\cite{agan2018ban, Lipton_DLP}.

Take the example of the U.S. ``Ban the Box'' campaign, which was aimed at passing legislation that required employers to be \emph{blind} to candidates’ criminal histories during initial assessments of qualifications\footnote{\url{https://bantheboxcampaign.org}}. The campaign was aimed at eliminating the check box on job applications that asked applicants to indicate whether they have a criminal history. Excluding criminal history from initial screenings of candidates captures formal EO's conception of a fair contest because it attempts to ensure that justice-involved persons are judged fairly on the basis of their qualifications and not dismissed out of hand. However, this \emph{formally fair} policy ended up having the opposite effect in practice.  Field studies showed that in the absence of individual information about applicants’ criminal histories, employers end up making group-level assumptions about prior criminal justice involvement ~\cite{agan2018ban}. This meant that applicants with no \eleni{justice involvement} who belonged to groups with higher (perceived) conviction rates, such as young black males, were adversely affected, while white applicants with \eleni{criminal justice involvement received the benefit of the doubt}. 

Similar statistical discrimination due to the exclusion of group-level information is seen when formal EO is encoded into algorithmic decision-making systems.  For example, \citet{Lipton_DLP} demonstrate a ``gender-blind'' algorithm that discriminates on the basis of ``inferred'' gender at the group level when gender information at the individual level is excluded.  As a result, the algorithm adversely treats applicants that it perceives as women (including men with long hair) and favors candidates that it infers to be men (including women with short hair).

The limitations of fairness through blindness are well-appreciated in fair-ML ~\cite{dwork_awareness}, so we instead turn to an alternative, stronger conception of formal EO.

\subsection{Formal EO as Calibration}
A well-calibrated test satisfies formal EO's conception of a fair contest because it ensures that the likelihood of getting a positive outcome does not depend upon morally arbitrary group membership ($s \in S$):
$$ P(y=1 | y'=c, s=0) = P(y=1 | y'=c, s=1).$$
Put differently, if two individuals have the same predicted score $y'$ (relevant merit) and only differ on group membership $s$ (morally irrelevant factors) then they are likely to get the same outcome from a well-calibrated test.

\subsection{Formal EO as Predictive Parity}
A test that satisfies predictive parity at threshold $p$ is formal-EO compliant because it ensures that the likelihood of getting a positive outcome is the same for all high-performing individuals, irrespective of morally arbitrary group membership ($s \in S$):
$$ P(y=1 | y'>p, s=0) = P(y=1 | y'>p, s=1).$$
Intuitively, formal EO mandates that all people who have job-relevant  qualifications ($y'>p$) should \eleni{ have the same chance of receiving} a positive outcome ($y=1$), irrespective of their \eleni{irrelevant,} morally arbitrary attributes ($s$), and predictive value parity as a fairness criterion reflects exactly this.

%% file: formal_plus.tex
\section{Formal-Plus EO}
\label{sec:formal_plus}
\rev{The strength of formal EO as a moral framework to design fair contests relies greatly on the ability to correctly measure candidates' relevant merit. In the codifications of formal EO as calibration and predictive parity, we are making an assumption about the predicted score $y'$, namely, that it does, in fact, measure the applicant's ``relevant merit''. This is a strong assumption and one that does not hold in societies with historic systemic inequality. \citet{Fishkin2014Bottlenecks} writes: ``When the formal egalitarian argued that the warrior children have more merit than the non-warrior children, that view depended on a factual premise: that the warrior test did what it was designed to do and accurately predicted future warrior performance. What if it did not?''}

\begin{figure}
\includegraphics[width=11cm]{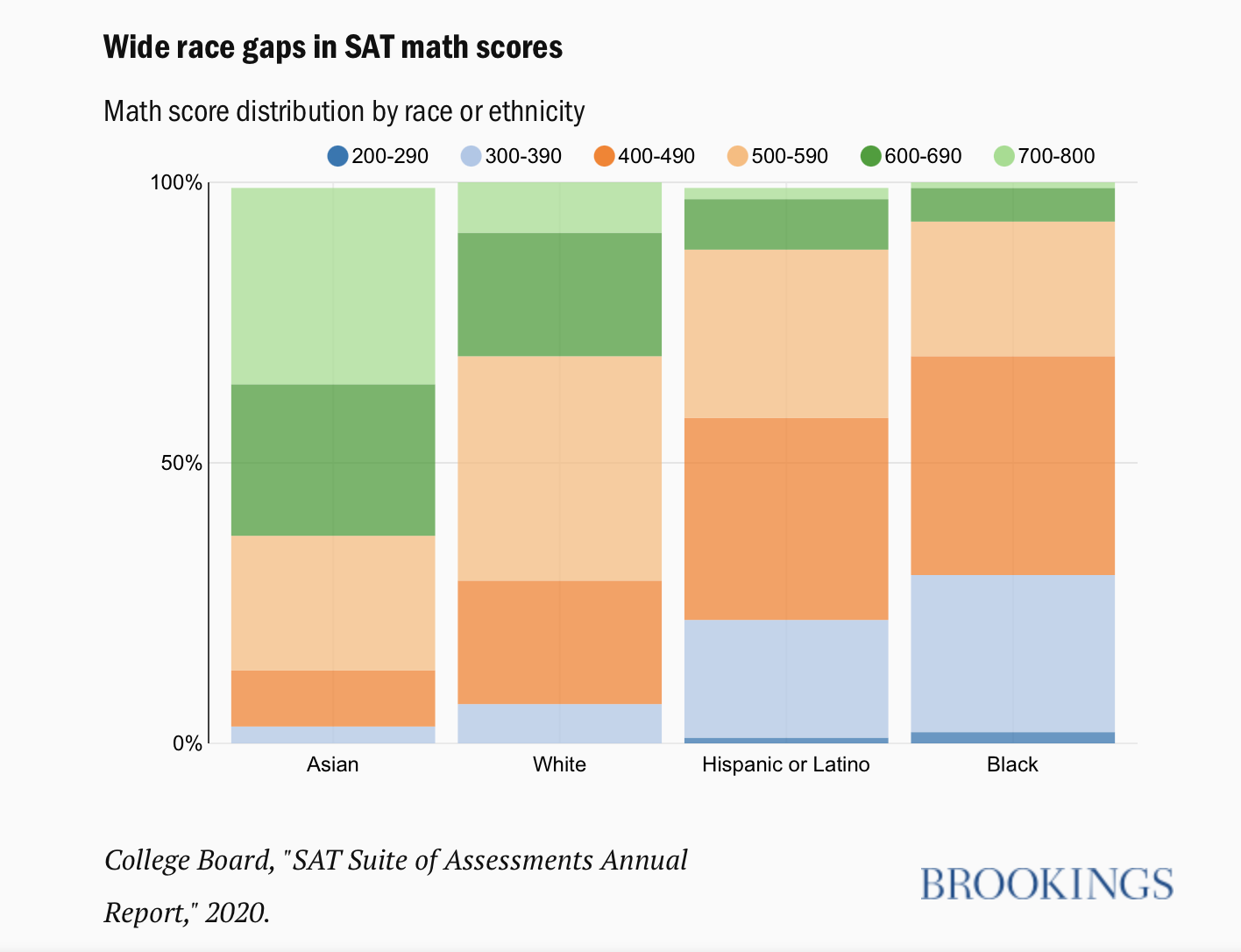}
\hfill
\caption{Distribution of SAT math scores by race or ethnicity}
\label{fig:SAT}
\end{figure}

For example, think of the SAT as a predictor of college success: When students can afford to do a lot of preparation, scores are an inflated reflection of students’ college potential. When students don’t have access to preparatory material, the SAT underestimates students’ college potential. The SAT systematically over-predicts the future performance of more privileged students, while systematically under-predicting future performance of less privileged students: that is, the test’s validity as a predictor of college potential varies across groups. Such disparity in standardized test scores along the lines of race (and gender) has been observed in real-world contexts and is shown in Figure \ref{fig:SAT}, reproduced from a Brookings 2020 report.\footnote{\url{https://www.brookings.edu/blog/up-front/2020/12/01/sat-math-scores-mirror-and-maintain-racial-inequity/}}

In order to correct for this circular problem, where the measurement of relevant merit itself tracks morally irrelevant privilege and disprivilege, \citet{Fishkin2014Bottlenecks} proposes a version of formal EO that he calls ``formal-plus''. \eleni{Formal-plus adjusts test results for members of groups that are systematically underestimated by a test. It thus} focuses on ensuring that test performance does not skew along the lines of morally irrelevant factors. From a formal-plus perspective, a \emph{fair contest} is one in which test errors do not track (morally arbitrary) group membership.

\subsection{Formal-plus EO as Error Rate Balance}
A test with balanced error rates at a threshold $p$ captures formal-plus EO's conception of a fair contest because it ensures that test performance (\ie false-positive rate and false-negative rate) does not skew with morally irrelevant group membership ($s \in S$): 
$$ P(y'>p | y=0, s=0) = P(y'>p | y=0, s=1) \text{ and}$$ 
$$ P(y'<=p | y=1, s=0) = P(y'<=p | y=1, s=1) $$

\subsection{Formal-plus EO as Equalized Odds}
\citet{Fishkin2014Bottlenecks} further explains that, in the absence of a ``perfectly accurate test'', and with the understanding of which groups the test tends to underestimate, the formal-plus EO conception of a \emph{fair contest} would ``give compensatory bonus points'' on the test to those whose future performance the test itself predictably underestimates. %\dots 
The idea of compensatory bonus points is simply to ``make more accurate predictions about who, in the future, will actually be the best warriors.'' The ``equal opportunity'' measure and algorithm from ~\citet{hardt_EOP2016} exactly captures formal-plus EO's moral desiderata, because it measures unfairness as the disparity in true-positive rates between groups, and randomly assigns positive outcomes when predicted scores fall between group-specific thresholds, in order to achieve the desired parity.

%% file: impossibility.tex
\section{Impossibility Results: Fair contests without fair life chances}
\label{sec:impossibility}

Concurrent work by \citet{chouldechova_impossibility} and 
\citet{Kleinberg_impossibility} showed that it is impossible to simultaneously achieve parity in error rates and positive predictive value for different groups, if the prevalence (or base rates) differs between these groups. We will now provide a moral interpretation of these impossibility results, through the lens of the EO doctrines that we have discussed so far, as the incompatibility between two different conceptions of a \emph{fair contest} --- one that seeks to fairly reward \emph{past} performance, versus one that seeks to fairly estimate \emph{future} performance --- when people do not have \emph{fair life chances}. We use this result to motivate the need for substantive conceptions of algorithmic fairness, which we discuss in the rest of the paper.

Formal EO's conception of a fair contest takes a moral \emph{end-point} view ~\cite{jencks} --- one that rewards the \eleni{qualifications} that people have \eleni{already developed}. Formal EO codified as predictive parity mandates that whoever possesses the relevant qualification score $y'>p$ (estimated from past performance) should be given the positive outcome $y=1$, irrespective of morally arbitrary group membership $s$. \citet{Kleinberg_impossibility}'s discussion of calibration as a fairness criterion is strikingly similar to the moral desiderata of formal EO. They write: ``This [calibration] means we are justified in treating people with the same score comparably with respect to the outcome, rather than treating people with the same score differently based on the group they belong to.'' This is exactly an end-point view of a fair contest --- one that rewards people with a comparable qualification score (computed on \eleni{existing merit, judged by} past performance) comparably. 

On the other hand, formal-plus EO's conception of a fair contest takes a humane \emph{starting-point} view ~\cite{jencks} --- one that wants to correctly estimate people's \eleni{likelihood of succeeding in the position on offer}. Once again, \citet{Kleinberg_impossibility}'s discussion of balance for the positive and negative class as a fairness criterion is strikingly similar to 
\citet{Fishkin2014Bottlenecks}'s conception of formal-plus EO: ``The second [balance for the negative class] and the third [balance for the positive class] ask that if two individuals in different groups exhibit comparable \emph{future behavior} (negative or positive), they should be treated comparably by the procedure.'' This is exactly formal-plus EO's starting-point view of a fair contest --- one that treats people with comparable estimated future performance comparably. Conversely, formal-plus EO would object to a test with unequal error rates across groups on the basis that the estimates of people's future performance skew along the lines of morally irrelevant privilege/disprivilege. This is also mirrored in the discussion from \citet{Kleinberg_impossibility}: ``In other words, a violation of, say, the second condition [balance for the negative class] would correspond to the members of the negative class in one group receiving consistently higher scores than the members of the negative class in the other group, despite the fact that the members of the negative class in the higher-scoring group have done nothing to warrant these higher scores.''

The prevalence or base rates in a particular population is the fraction of people who possess a certain quality/qualification or receive a positive outcome (based on the existence of that quality) \cite{chouldechova_impossibility, Kleinberg_impossibility}. For example, in the context of hiring, the base rate in the female population is the fraction of female candidates who receive a positive outcome --- a hiring offer --- among all the female applicants. Fair life chances is a broad philosophical concept, but in the context of a discrete decision, we can think of base rates among populations as a proxy for their life chances. For example, if there was gender-equality in the workforce, and women had the same employment prospects as men, then we would expect equal proportions of women and men to receive a positive hiring outcome. 

Putting this together, a philosophical interpretation of the impossibility results, through the lens of EO doctrines, says that it is impossible to design a \emph{fair contest} that simultaneously rewards people's past qualifications and accurately estimates people's future prospects of success, if people did not have fair (comparable) life chances. This is simply because morally arbitrary and irrelevant factors weigh heavily on people's achievement --- both as evidenced in past performance, and in the estimation of future performance.

The empirical results in critical domains such as criminal justice that led to the impossibility results in fair-ML \cite{chouldechova_frontiers, Kleinberg_impossibility} are a stark demonstration of this fact: that people do not have comparable life chances and that morally arbitrary characteristics such as gender and race do weigh heavily on people's prospects of success. Importantly, through the lens of EO doctrines, we can see the limitations of our current approaches in their narrow focus on designing fair contests at discrete decision points. We now discuss substantive EO doctrines, to pivot future directions of fair-ML research towards substantive conceptions of algorithmic fairness.

%% file: substantive.tex
\section{Substantive EO}
\label{sec:substantive}

\rev{In order to design interventions that improve people's life chances we need substantive EO.} Substantive EO focuses on giving people the opportunities to substantively build up their qualifications, so that when they do go on to compete for desirable social positions, they truly have a chance of winning. These qualifications-building opportunities fall into the second domain of EO, discussed in Section~\ref{sec:eo:domains}, and constitute \emph{developmental opportunities}. The motivating reason behind both equality of developmental opportunities and EO over a lifetime is that morally arbitrary circumstances of birth, such as warrior parentage, should not determine people's \emph{life} prospects.

There are several conceptions of substantive EO, but in this paper we will limit our discussion to two highly influential doctrines that are relevant to fair-ML, namely, \rev{Rawls's fair EO ~\cite{Rawls1971Justice} and luck-egalitarian EO ~\cite{dworkin_1981, Roemer2002}.}

\subsection{Limitations of formal doctrines}
\rev{Before we move on to substantive EO, let us look at why formal EO doctrines fall short of satisfying all of our fairness-related concerns. Returning momentarily to Bernard Williams’s famous example of a warrior society \cite{williams_1973}: The battle may be formally \emph{fair}, in that everyone is allowed to compete \eleni{and competitors are judged only on the basis of relevant characteristics}, but the deck is stacked in favor of the children of warriors. Arbitrary and morally irrelevant privileges weigh heavily on the outcomes of formally \emph{fair} competitions because people can leverage them to build qualifications in advance of competitions. This undermines the promise of formal doctrines --- that only relevant skill will be rewarded --- because it fails to stop gains from being distributed along the lines of privilege and disprivilege. We call this the \emph{before} problem: before competitions, people are allowed to exercise their privilege to develop relevant qualifications.}

\rev{Formal doctrines also have an ``after'' problem.  After formally \emph{fair} competitions, winners are set up for even more success.  A candidate that is hired for a job is consequently granted access to more training and job experience. This makes them even more competitive in the next competition for jobs. \eleni{Conversely, those who lose at first, lose opportunities for skill development, leading to more losses. Formally fair competitions create a snowball effect, where early-on winners get further ahead while early-on losers fall further behind.}}

\rev{Formal EO's ``before'' and ``after'' problems compound,} \eleni{with privileged candidates using their competitive advantages to win early on, thus securing more developmental opportunities, which enable further wins. Worse yet, if an early competition is not formally fair---say, a candidate is awarded a job in part due to family connections---that candidate can take advantage of consequent on-the-job training to develop qualifications that help them win future job competitions fair and square.} \citet{anderson_integration} calls this phenomenon ``discrimination laundering'': ``while discrimination in hiring and on-the-job training is illegal, discrimination on the basis of differences in human capital due to differences in on-the-job training is not.'' {Formal doctrines cannot adequately address these problems, because they are limited to measuring people’s qualifications accurately, and to excluding irrelevant information.} 

\rev{To correct for social inequalities in any meaningful way we need substantive EO.} \citet{Fishkin2014Bottlenecks} writes: ``The reason that the warrior society is interesting is that, per stipulation, it is not simply the case that children of warriors appear, through test-related artifice, most likely to be the best future warriors. The point of the example is that the children of warriors really are the most likely to grow into the best adult warriors as a result of their accumulated childhood advantages.''

This is exactly the target of substantive EO doctrines: making sure that people have comparable ``opportunity sets'' over the course of a lifetime.  

\input{rawls}

\input{luck}

\section{A modern re-interpretation of substantive EO doctrines}
\label{sec:substantive-modern}

Substantive EO doctrines discussed in Section \ref{sec:substantive} have much stronger moral desiderata than formal EO doctrines discussed in Sections \ref{sec:formal} and \ref{sec:formal_plus}, but they also have severe shortcomings that can preclude applying them to real-world contexts that are the focus of fair-ML. Rawls' theory has been the subject of much debate and criticism for being limited to ideal theorizing: There is very little guidance from Rawls about how to apply his principles of justice in practice, or how to bring about FEO in a world where people do not have comparable life prospects. Luck-egalitarianism, specially strict interpretations of the doctrine, also has severe shortcomings such as issues of agency and autonomy in holding people responsible for certain types of luck and not for others. 

In this section we provide a modern interpretation of luck-egalitarian EO and Rawls' FEO, in a way that is both consistent with the original doctrines, and amenable to providing normative guidance in practical contexts: We classify the luck-egalitarian approach as a \emph{backward-facing}, indirect approach to equalizing people's opportunity sets --- a luck-egalitarian conception of a \emph{fair contest} is one where outcomes are distributed after correcting for unequal life chances in the \emph{past}. By contrast, we classify Rawls as a \emph{forward-facing}, direct approach to bringing about substantive EO --- a Rawlsian conception of a \emph{fair contest} optimizes for equality in the \emph{future} life chances of equally talented people.  

Table~\ref{table:classification} summarizes our classification of the normative approaches of different EO doctrines.  We will elaborate on this in the remainder of the section.

\begin{table}[h!]
\caption{Classification of EO doctrines}
\centering
\begin{tabular}{||c| c |c ||} 
 \hline
 & Backward-facing & Forward-facing \\ [0.5ex] 
 \hline\hline
Fair contests & Formal & Formal-plus \\ 
 \hline\hline
Fair life chances & Luck-egalitarian & Rawls \\
 \hline
\end{tabular}
\label{table:classification}
\end{table}

\subsection{Luck-egalitarian EO as a \emph{backward-facing} view of Fair Life Chances}
The luck-egalitarian view acknowledges that differences in people’s qualifications at the point of competition are, at least in part, due to morally arbitrary circumstances (matters of brute luck) and so a fair competition should only evaluate candidates on the basis of their propensity to expend effort, and not on the qualifications that are built from this effort. Intuitively, the idea is that people who are disadvantaged by circumstances will have to put in far greater effort to reach the same level of ability as compared to people with advantageous circumstances, and so effort is the correct rubric of achievement, not ability.

From a practical standpoint, the luck-egalitarian approach gives rise to a two-step procedure of substantive EO: first, control for people’s unequal life chances, and then conduct fair contests on the basis of these adjusted qualifications. We interpret luck-egalitarianism as a \emph{backward-facing} conception: it corrects for \emph{past} effects of brute luck, and then allows individual effort to decide future outcomes. It does improve people's life chances by distributing opportunities to which they likely would not have had access, had we not corrected for their unequal life chances in the past.  However, this conception does not correct for the differential effort that will be required by people with different circumstances to excel in that position in the \emph{future}. 

\subsection{Rawls' FEO as a \emph{forward-facing} view of Fair Life Chances}

Rawls' principle of fair equality of opportunity (FEO) says that \emph{equally talented people should have equal prospects of success}. Let us unpack the principle into two components: the first part deals with identifying ``equally talented'' people, whereas the second part says that outcomes should be distributed in such a way that gives these ``equally talented'' people  ``equal prospects of success''.

An implementation of the first part of the principle aligns with the luck-egalitarian approach: we would have to adjust our measurement of people's abilities after controlling for the effects of the social lottery in order to approximate their ``native talent''. The second part of the principle is where the two conceptions diverge: the luck-egalitarian stops after correcting for \emph{past} effects and simply distributes outcomes based on this corrected measurement. Rawls goes one step further, and distributes outcomes in a way that also makes people's \emph{future} prospects of success (\ie their prospects of succeeding in the next contest) comparable. Hence, we interpret Rawl's FEO as a \emph{forward-facing} view of the principle of fair life chances.

%A summary of our classification of the normative approaches of different EO doctrines is given in Table \ref{table:classification}. 
We summarize our interpretation of EO doctrines in the next section, and go on to illustrate the distinction between practical applications of different EO doctrines with the help of examples in Section~\ref{sec:normative}.

%% file: rawls.tex
\subsection{Rawlsian Fair EO}
\label{sec:rawls}
Rawls' principle of fair equality of opportunity (FEO) says that \emph{equally talented people should have equal prospects of success}. \citet{Rawls1971Justice} writes: ``Assuming that there is a distribution of natural assets, those who are at the same level of talent and ability, and have the same willingness to use them, should have the same prospects of success regardless of their initial place in the social system.''
In setting out his theory of justice, in which the principle of FEO is embedded, Rawls identifies two distributive mechanisms: the ``social lottery'' and the ``natural lottery''. The social lottery distributes people their initial positions in the social system, whereas the natural lottery distributes people their native talent and ability. Rawls characterizes both as morally arbitrary. He writes ~\cite{Rawls1971Justice}: ``We do not deserve our place in the distribution of native endowments, any more than we deserve our initial starting place in society.'' Rawls posits that the natural and social lotteries are not by themselves unjust, but it is the way that institutions have been set up that leads to inequality along the lines of morally arbitrary characteristics. His principles of justice are designed to help a society appropriately mitigate their effects. He writes ~\cite{Rawls1971Justice}: ``The natural distribution is neither just nor unjust; nor is it unjust that persons are born into society at some particular position. These are simply natural facts. What is just and unjust is the way that institutions deal with these facts.'' 
%Instead of positing the social and natural lotteries as a source of \emph{past} injustice that must be corrected for, Rawls takes a \emph{forward-facing} view that instead regulates the distribution of primary social goods (such as wealth and opportunity) by institutions. 
With this in mind, Rawls' theory of justice \cite{Rawls1971Justice} posits the following principles that would regulate the distribution of primary social goods (including wealth and opportunity) by institutions in a just society:
\begin{enumerate} 
    \item~Rights and liberties: Everyone has the same inalienable right to equal basic liberties.
    \item
    \begin{enumerate}
        \item~Principle of Fair EO: All offices and positions must be open to all under conditions of fair equality of opportunity.
        \item~Difference principle: Any social inequality must be applied in such a manner that they be of the greatest benefit to the least advantaged.
    \end{enumerate}
\end{enumerate}

The principles are lexically ordered, in that people's basic rights and liberties cannot be infringed upon while bringing about FEO, nor can the Difference Principle be applied in a way that violates FEO. People's fundamental rights are given highest priority. Next, the principle of FEO regulates the distribution of desirable social positions such that people benefit from their arbitrary endowments from the natural lottery, but are not disadvantaged by the social lottery. Once this has been satisfied, the Difference Principle is applied, seeking to redistribute social inequality to the greatest benefit of the worst-off group, so as to limit the current and \emph{future} effect of both lotteries.

Rawls' principles, including his principle of fair EO, are regulatory and holistic in nature --- they are to be applied iteratively to regulate the distribution of social goods by institutions. It is unclear how to port these principles, as currently understood, to discrete decision-making contexts ---  the kind that are the focus of fair-ML. In Section \ref{sec:substantive-modern} we will propose a modern re-interpretation that allows for such a translation.

%% file: luck.tex
\subsection{Luck Egalitarian EO}
\label{sec:luck}
Luck-egalitarian EO levels the playing field by making competitors’ opportunities comparable, and then allows individual choices and effort to determine the outcomes of competitions. Any resulting disparity in outcomes is morally acceptable because it is due to differential individual effort, not differential fortune.

At a discrete decision point (the first domain of EO), morally arbitrary circumstances have already weighed heavily on people’s abilities. The luck-egalitarian conception of a fair contest partitions a person’s qualifications into two sets --- matters of ``option luck'' or ``choice luck'' for which it is morally correct to hold the individual accountable, and effects of ``brute luck'' that are morally irrelevant. Luck-egalitarian EO says that people’s outcomes (access to desirable positions) should only be affected by the former, and no matters of brute luck should affect the outcome of a fair contest.

The hard question now is how to make this correction. How do we separate the effects of brute luck (circumstance) from the effects of responsible choices (effort)? \citet{Roemer2002} proposed a version of EO that fulfills the moral desiderata of the luck-egalitarian doctrine while bypassing the need to make an explicit separation between ``responsible effort'' and ``arbitrary circumstance''. Instead, Roemer introduced the idea of ``types'': people with the same morally arbitrary circumstance are of the same ``type''. Now, for a certain matter of arbitrary circumstance (\eg family income), the entire population can be partitioned into types (\eg ``high income'', ``medium income'', and ``low income''). Using this idea of circumstance-types, Roemer posits that, in comparing the effort of candidates, we should correct for the fact that those efforts are \emph{drawn from different distributions}.  In other words, effort distributions are characteristic of the type, and not of the individual, and this difference is due to a morally arbitrary factor of circumstance, for which individuals should not be held accountable. Now, in evaluating an individual’s qualifications (effort), we should only compare them to others of the same type (with the same circumstance). For example, in Figure \ref{fig:SAT}, we see a large disparity in the SAT score distributions broken down by race or ethnicity. In a luck-egalitarian procedure, we would evaluate students' test scores based on where they placed within the score distribution of their particular race/ethnicity. From a moral standpoint, two individuals of different types are equally qualified for a desirable position if they lie at the same quantile of the effort distribution of their type. Hence, Roemer’s conception of the a \emph{fair contest} evaluates people by correcting for the unequal (unfair) life chances they've had in the past by ranking them in their effort-type distribution.

%% file: taxonomy.tex
\section{Fairness as Equal Opportunity Taxonomy}
\label{sec:taxonomy}

We summarize the moral desiderata and normative approaches of different EO doctrines in the \emph{Fairness as Equal Opportunity} taxonomy given in Table~\ref{table:normative}. This gives us guidance about what value judgements our fairness interventions codify, and helps us design a suitable fairness intervention for a given context based on our normative judgements.  For example, we can decide that our outcomes correspond to rewards for past performance and choose a \emph{formal} approach. Or, we can decide that the decision-making context requires selecting people who are most likely to succeed in the future and take a \emph{formal-plus} approach instead. We can apply a two-step substantive approach following the luck-egalitarian view by first adjusting the measurement of people's qualifications to correct for past effects of morally arbitrary factors, and then apply any suitable selection procedure on these adjusted qualification scores. Lastly, we can gauge that the decision-making context is a critical developmental opportunity, and choose a Rawlsian approach that forgoes maximum utility today in favor of improved equity tomorrow.  

\begin{table}[h!]
\caption{Fairness as Equal Opportunity taxonomy}
\centering
\begin{tabular}{||c| c |c ||}
\hline
Doctrine & Moral desiderata & Normative approach \\ [0.5ex] 
\hline\hline
Formal &  \makecell{Fair contests should only measure \\ morally relevant qualifications} & Accurately measure past performance \\ 
\hline\hline
Formal-plus & \makecell{The performance of fair contests \\ should not skew along the lines of \\ morally irrelevant features} & Accurately estimate future performance\\ 
\hline\hline
\makecell{Substantive: \\ Luck-egalitarian} & \makecell{Matters of brute luck should \\ not affect people's outcomes} & \makecell{Distribute outcomes on the basis of effort, \\ after correcting for the past effects \\  of morally arbitrary circumstances} \\ 
\hline\hline
\makecell{Substantive: \\ Rawls} & \makecell{Equally talented people should \\ have equal prospects of success} & \makecell{Distribute outcomes to equalize future \\  prospects of success of people who have the same \\ native talent, irrespective of arbitrary circumstance} \\
\hline
\end{tabular}
\label{table:normative}
\end{table}

%% file: normative.tex
\section{Normative Guidance}
\label{sec:normative}

\subsection{An Illustrative Example: College Admissions}
\label{sec:example}

\begin{figure}
\includegraphics[width=12cm]{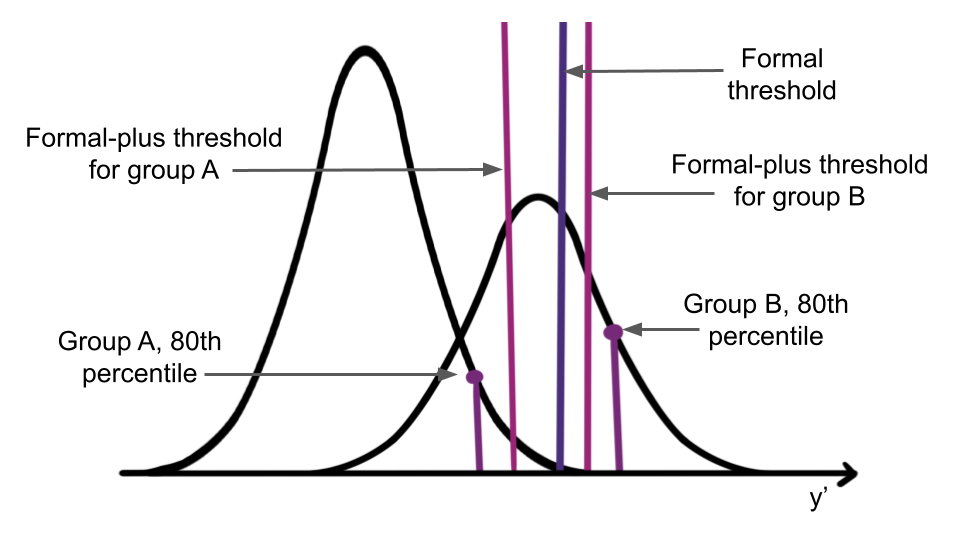}
\hfill
\caption{Distribution of test scores for different demographic groups in the college admission example discussed in section \ref{sec:example}}.
\label{fig:example}
\end{figure}

We now present an example to illustrate how different EO doctrines conceptualize a \emph{fair contest}. Suppose, for the sake of argument, that we are making a college admissions decision, based only on a single standardized score, shown in Figure~\ref{fig:example}. We have two demographic groups A and B, where group membership is based on some morally arbitrary characteristic, say race or gender. We can see that members of group A (the distribution to the left) have systematically lower scores than members of group B (the curve to the right), and we posit that this is due to the effect of morally arbitrary circumstances that differ between the two groups, and are not due to any innate difference in talents in the two groups. The question now is: How do we distribute the desirable outcome of a positive admissions decision to people from both groups, in a way that is \emph{fair}?

Following formal EO, we would simply decide a threshold on the score, which corresponds to the level of past performance that we find deserving of a reward. This threshold is shown in the figure as ``Formal threshold''. Positive outcomes are distributed to everyone who has a score ($y'$) that is higher than this threshold. As we can see from the figure, although we did not explicitly set this threshold based on any morally irrelevant factors (we simply decided a cut-off on the standardized test score), nonetheless, this threshold is prohibitively high and our selection procedure effectively eliminates all of group A. This is a backward-facing view of a \emph{fair contest}: one that distributes outcomes based on people's past performance on the standardized test.  

A formal-plus conception of this procedure would posit that the standardized test overestimates the abilities of group B and underestimates  the abilities of group A, and so the test is not a suitable measurement of applicants' \emph{future} academic performance. Alternatively, it would select different thresholds for each group, shown in the figure, to correct for the test's error. The formal-plus threshold for group A does admit some people from this group, and this is exactly Fishkin's idea of ``compensatory bonus points'' for the group whose abilities the test underestimates. This is a forward-facing view of \emph{fair contests}: one that distributes outcomes based on accurately estimating people's future performance.  

Next, a luck-egalitarian approach following Roemer would look at people's positions within the score distributions of their type (the 80th percentile for group A and group B are shown in Figure \ref{fig:example}). The luck-egalitarian would posit that people who are at the same percentile in the score distribution of their type have expended the same degree of effort in the past, and hence should receive similar outcomes from the procedure. This is a backward-facing view of substantive EO: one that designs fair contests by correcting for unequal life chances in the past, and adjusts people's qualification score to only reflect their morally relevant effort. 

In order to differentiate the Rawlsian approach to this problem from the luck-egalitarian one, let us further look at two individuals, Alice (from type A) and Bob (from type B), who sit at the same percentile of the score distribution of their type, and hence are equally ``talented'' according to the Rawlsian view. We saw that the luck-egalitarian would give them both the same outcome (\ie a positive admission decision) because these individuals have demonstrated the same degree of effort in the past. By contrast, the Rawlsian would make a further consideration: How likely are Alice and Bob to succeed in this desirable position, respectively? Even if they both receive the positive outcome, Alice will probably have to work much harder than Bob to actually do well in the program. The effects of circumstance are such that the absolute amount of effort they have to spend to match the same level of achievement is higher for Alice than for Bob. 

The luck-egalitarian view does not correct for the continued future effects of circumstance, while the Rawlsian approach tries to, and makes positive admissions decisions for both Alice and Bob, but also distributes additional resources (such as tutoring and scholarships) for Alice to make up for the lack of resources and developmental opportunities they had access to, leading up to this competition. This is why we classify Rawls' FEO as a forward-facing view of a \emph{fair contest}: it wants to set up equally talented people to have equal prospects of winning the next contest.

Note that the actual outcomes distributed by these different procedures (motivated by different EO doctrines) may, in fact, coincide in certain contexts, based on the relative effects of morally irrelevant features. What distinguishes these procedures are their normative approaches and value judgements, and not the outcomes that are finally distributed. 

\subsection{An illustrative example: COMPAS}
\label{sec:contexts}

We now demonstrate how the taxonomy of EO doctrines in Table~\ref{table:normative} enables us to debate in normative terms, based on our value judgements. Let us consider the infamous example of COMPAS, as exposed in an investigation by ProPublica~\cite{compas_propublica}. \rev{We would like to clarify that we chose COMPAS as the example with which to illustrate our framework not because we believe it to be representative of the kinds of contexts for which algorithms \emph{should} be designed, and \emph{could} benefit from ethical and moral grounding. On the contrary, we unequivocally believe that an algorithm such as COMPAS \emph{should not} be used. %However, the truth is that it is used, and, furthermore, some technologists reject ProPublica's audit of COMPAS and uphold that COMPAS is well-calibrated and thereby non-discriminatory~\cite{}. 
That being said, without grounding different statistical measures of fairness in the moral desiderata they encode we have no way to reconcile disagreements about the implications of ProPublica's findings. We view our EO-framework as a necessary first step towards being able to debate in values, and to audit algorithmic systems more holistically}.

\rev{COMPAS is an algorithm that predicts the risk of violent recidivism among \eleni{people awaiting trial}. This can be posed as the problem of distributing access to resources such as  counseling or other positive interventions, based on \eleni{invididuals'} propensity to re-offend, and translates to a problem to which we can apply EO doctrines. Northpointe argued that COMPAS did not exhibit racial discrimination because the risk scores that it produced were equally well calibrated for both black and white defendants ~\cite{northpoine}. Connecting this argument to the moral desiderata of formal EO, we see that Northpointe's value judgement was that the algorithm ``accurately'' rewarded (and punished) people for their \emph{past} actions (\ie past criminal behavior). ProPublica, on the other hand, demonstrated that the error rates of COMPAS skewed along racial lines --- the algorithm systematically overestimated the risk of black \eleni{individuals} and systematically underpredicted the risk of white \eleni{individuals} --- and argued that this was evidence of racial discrimination ~\cite{compas_propublica}. ProPublica's critique of COMPAS can be seen through the lens of formal-plus EO: the use of COMPAS to determine parole sentences is not formal-plus EO compliant because test performance skews along the lines of race --- a morally arbitrary and irrelevant feature. ProPublica \eleni{appears to be} making the value judgement that COMPAS ought to accurately estimate the \emph{future} prospects of crime of both black and white \eleni{individuals}. Given that the purpose of COMPAS was, in fact, to \emph{predict recidivism} (\ie criminal re-offense in the \emph{future}), viewing ProPublica's audit through the lens of formal-plus EO makes their results even more compelling. }

%% file: related_work.tex
\section{Related Work}
\label{sec:related_work}

\citet{heidari_EOP} were the first to apply ideas of Equal Opportunity to fair-ML, using economic models of EO. We were inspired by their work, but, in constructing this taxonomy using EO doctrines from political philosophy, find critical mistakes in their framework. Most importantly, their framework takes a reductive view of substantive EO doctrines, as independent fairness criteria at discrete decision points. This misrepresents the nature of substantive EO, which is not limited to the first domain of EO and takes a more holistic view of \emph{fair contests} in regards to people's \emph{fair life chances}, as we explain in Sections~\ref{sec:eo} and~\ref{sec:substantive}. Substantive EO doctrines are broad guiding procedures that cannot be collapsed into a single statistical measure. This is not a limitation, but rather the nature of these doctrines --- they do not apply to a single decision point, but to opportunity sets distributed over time.

With this clarification, we refute \citet{heidari_EOP}'s mappings of substantive EO doctrines with statistical measures, including their assertion that statistical parity, equalized odds, and accuracy map to Rawlsian EO. We agree with their characterization of Roemer's EO, and their application of it to predictive contexts. An important point here is that formal and Rawls' FEO are doctrines from political philosophy, and not economics, while Roemer's EO is an economic doctrine. Hence, \citet{heidari_EOP}'s choice to use economic models of EO might be the cause of disagreement between our framework (using EO doctrines from political philosophy) and theirs (using economic models of EO). 

Specifically, \citet{heidari_EOP} characterize methods that are consistent with luck-egaitarian EO as taking a relative view of effort, and they characterize methods that are consistent with Rawls' FEO as taking an absolute view of effort.  The latter is a misconception, in that 
%Strikingly, 
even economic interpretations of Rawls' FEO do not characterize it as a doctrine that takes an absolute view of effort~\cite{Lefranc_Trannoy}.

Further, following~\citet{Arneson2018FourCO}, libertarianism has been introduced as a possible version of EO in fair-ML~\cite{heidari_EOP}. However, the libertarian view focuses on a narrow notion of procedural fairness: It would object to a procedure that allows illegal or unfair means of gaining access to opportunities. While libertarianism (as a limited notion of procedural fairness) may be interpreted as a fairness-preserving position from a legal standpoint, it does not satisfy EO’s characteristic commitment to eliminating irrelevant and arbitrary barriers to achievement. The libertarian principle of self-ownership asserts that people are entitled to the full benefit of their natural personal endowments~\cite{Lefranc_Trannoy}, and so the disparity in people's access to desirable positions arises simply from them exercising free will. According to this view, nothing is morally arbitrary or irrelevant, and needs to be corrected for. Hence, we reject \citet{Arneson2018FourCO}'s characterization of liberatarianism as a form of EO, and exclude it from our fairness-as-EO framework.

There has been some interest in fair-ML to move beyond a single-point view of algorithmic fairness: \citet{Liu2021Rawlsnet} developed RawlsNet, which uses a Bayesian network to adjust people's qualification score by modifying the outcomes of previous competitions, such as competitions that determine access to educational opportunities. This approach attempts to explicitly codify Rawls's FEO principle; it certainly corrects for the social lottery and, towards that end, is complementary to the Rawlsian substantive procedure described by us. It differs from our framework in that it does not have a forward-facing component that also seeks to improve parity in people’s future prospects of success, and instead takes a backward-facing approach by conducting fair contests after creating equality in developmental opportunities.

A particularly relevant line of work in fair-ML is from \citet{liu2018delayed}, on the delayed impact of fair decision-making on people’s prospects of success in the future. While they do not cite EO doctrines as the basis for taking this approach, this procedure certainly has a substantive flavor: it looks at the effect of over/under-selecting candidates from demographic groups without substantively altering their competitive qualifications, thereby measuring the (delayed) impact of conducting fair contests without improving people’s life chances.

Another relevant line of work is from \citet{downstream_kannan} and \citet{multistage-Facct22}, on incorporating Equal Opportunity objectives in multi-stage decision-making pipelines. \citet{downstream_kannan} design fairness criteria based on candidates' ``types''. Again, while they do not cite any EO doctrines from political philosophy as the inspiration behind their model, their equal opportunity criterion strongly resembles Roemer's conception of luck-egalitarian EO \cite{Roemer2002}. \citet{multistage-Facct22}, on the other hand, use \citet{hardt_EOP2016}'s equal opportunity measure, which, as we discussed in Section \ref{sec:formal_plus}, maps to the formal-plus doctrine, and is not a substantive conception of EO.

To the best of our knowledge, we are the first to arrange fairness desiderata under the two EO principles of \emph{fair contests} and \emph{fair life chances}, and to introduce \citet{Fishkin2014Bottlenecks}'s conception of formal-plus EO as a criterion on error rate difference between groups to the fair-ML community, which is a critical component of our fairness-as-EO taxonomy. 

%% file: conclusion.tex
\section{Discussion}
\label{sec:discussion}

It is widely accepted that \emph{fairness} is not a statistical concept, but rather a philosophical \eleni{and moral} one. Yet, current approaches in fair-ML do not explicitly state the value judgements they make. In this work, we attempt to fix this deficit by making connections between influential results in algorithmic fairness, and the normative considerations of equal opportunity doctrines. Through a taxonomy, summarized in Table \ref{table:normative}, we can finally debate in value judgements and not limit our discussion to a purely technical or statistical methodology.

In making these connections between EO doctrines and algorithmic fairness approaches, we have identified limitations in current approaches, specifically, a narrow focus on designing \emph{fair contests} at discrete decision points, without broader considerations of \emph{fair life chances} and the overall opportunity-sets available to people. We also propose two broad methodologies for a substantive conception of algorithmic fairness, motivated by substantive doctrines of equal opportunity. 

The first approach, motivated by the luck-egalitarian perspective, follows a two-step procedure: First, it adjusts people's measured qualifications by correcting for effects of morally irrelevant circumstance. This is a backward-facing view, which acknowledges that people did not have \emph{fair life chances} in the past. The second step is to conduct a \emph{fair contest} in such a way that people who have demonstrated the same ability (propensity to expend effort) should receive comparable outcomes. We view this as an indirect approach to equalizing people's life chances, because it only corrects for past effects of morally arbitrary circumstances and does not distribute outcomes in anticipation of future effects. 

A second approach, motivated by Rawls's principle of fair equality of opportunity, is more direct: It acknowledges both that people's initial endowments themselves are arbitrary, and that the measurement of these talents in any competition is further affected by morally arbitrary initial social conditions. Given that people do not deserve the benefits nor the disadvantages of their social endowments, a Rawlsian approach to EO first measures native talent by removing the effects of the social lottery. Then, based on these native talents, it distributes outcomes in a way that equalizes the prospects of equally talented people. This is a forward-facing approach, because it designs \emph{fair contests} to set up equally talented people to have comparable prospects of winning the next \emph{fair contest}. 

It is also widely accepted that \emph{fairness} is inherently context-specific, yet, to the best of our knowledge, an understanding of the suitability of different fairness conceptions in different contexts still lacks. We attempt to take a first step in this direction: Equal opportunity doctrines, in contrast to equality of outcome doctrines, allow us to connect the nature of the opportunity with fairness desiderata. Our taxonomy makes explicit the value judgements that go into different conceptions of a \emph{fair contest}, underscoring that different conceptions are suitable for different contexts: Do we seek to reward past performance (formal) or accurately estimate future performance (formal-plus)? Do we care about removing the effects of past inequality (luck-egalitarian), or does our context require that we preemptively correct for inequality that, if left unchecked, will compound in the future (Rawls)? Thinking in these terms can guide decision-makers in selecting a suitable fairness-related intervention that aligns with their value judgements for what is suitable for their specific context. 

Lastly, we would like to re-iterate the limitations of this framework: EO doctrines are only applicable to distributive contexts, where desirable outcomes are distributed on the basis of some relevant qualification.

\section{Conclusion}
\label{sec:conclusion}

In this work we showed that extant approaches to algorithmic fairness have mainly been limited to formal conceptions of \emph{fair contests} at discrete decision points. Through the lens of EO doctrines, we provided a moral interpretation of the impossibility results as the incompatibility between two different conceptions of a \emph{fair contest} --- a forward-facing view versus backward-facing one --- when people do not have \emph{fair life chances}. We used this result to motivate the need for substantive conceptions of algorithmic fairness, which look more holistically at the opportunity sets that people have available to them over the course of a lifetime, and outlined two plausible procedures to do this. We hope that our work will foster similar approaches from law and social sciences in grounding current and future research in fair-ML in strong normative foundations.

%% file: main.bbl
\begin{thebibliography}{25}
\providecommand{\natexlab}[1]{#1}
\providecommand{\url}[1]{\texttt{#1}}
\expandafter\ifx\csname urlstyle\endcsname\relax
  \providecommand{\doi}[1]{doi: #1}\else
  \providecommand{\doi}{doi: \begingroup \urlstyle{rm}\Url}\fi

\bibitem[Anderson(2010)]{anderson_integration}
Elizabeth Anderson.
\newblock \emph{The Imperative of Integration}.
\newblock Princeton University Press, 2010.
\newblock ISBN 9780691139814.

\bibitem[Anderson(1999)]{Anderson1999Equality}
Elizabeth~S. Anderson.
\newblock What is the point of equality?
\newblock \emph{Ethics}, 109\penalty0 (2):\penalty0 287--337, 1999.
\newblock ISSN 00141704, 1539297X.
\newblock URL \url{http://www.jstor.org/stable/10.1086/233897}.

\bibitem[Arneson(2018)]{Arneson2018FourCO}
Richard~J. Arneson.
\newblock Four conceptions of equal opportunity.
\newblock \emph{Wiley-Blackwell: Economic Journal}, 2018.

\bibitem[Fishkin(2014)]{Fishkin2014Bottlenecks}
Joseph Fishkin.
\newblock \emph{Bottlenecks: A New Theory of Equal Opportunity}.
\newblock Oup Usa, 2014.

\bibitem[Dworkin(1981)]{dworkin_1981}
Ronald Dworkin.
\newblock What is equality? part 1: Equality of welfare.
\newblock \emph{Philosophy and Public Affairs}, 10\penalty0 (3):\penalty0
  185--246, 1981.
\newblock ISSN 00483915, 10884963.
\newblock URL \url{http://www.jstor.org/stable/2264894}.

\bibitem[Lefranc et~al.(2009)Lefranc, Pistolesi, and Trannoy]{Lefranc_Trannoy}
Arnaud Lefranc, Nicolas Pistolesi, and Alain Trannoy.
\newblock Equality of opportunity and luck: Definitions and testable
  conditions, with an application to income in france.
\newblock \emph{Journal of Public Economics}, 93\penalty0 (11):\penalty0
  1189--1207, 2009.
\newblock ISSN 0047-2727.
\newblock \doi{https://doi.org/10.1016/j.jpubeco.2009.07.008}.
\newblock URL
  \url{https://www.sciencedirect.com/science/article/pii/S0047272709000905}.

\bibitem[Rawls(1971)]{Rawls1971Justice}
John Rawls.
\newblock \emph{A Theory of Justice}.
\newblock Harvard University Press, 1971.
\newblock ISBN 9780674880108.
\newblock URL \url{http://www.jstor.org/stable/j.ctvjf9z6v}.

\bibitem[Roemer(2002)]{Roemer2002}
John Roemer.
\newblock Equality of opportunity: A progress report.
\newblock \emph{Social Choice and Welfare}, 19\penalty0 (2):\penalty0 455--471,
  2002.
\newblock URL
  \url{https://EconPapers.repec.org/RePEc:spr:sochwe:v:19:y:2002:i:2:p:455-471}.

\bibitem[Williams(1973)]{williams_1973}
Bernard Williams.
\newblock \emph{The idea of equality}, page 230–249.
\newblock Cambridge University Press, 1973.
\newblock \doi{10.1017/CBO9780511621253.016}.

\bibitem[Young and Foundation(1994)]{young1994equity}
H.P. Young and Russell~Sage Foundation.
\newblock \emph{Equity: In Theory and Practice}.
\newblock A Russell Sage Foundation book. Princeton University Press, 1994.
\newblock ISBN 9780691043197.
\newblock URL \url{https://books.google.co.in/books?id=XVK5AAAAIAAJ}.

\bibitem[Jencks(1988)]{jencks}
Christopher Jencks.
\newblock Whom must we treat equally for educational opportunity to be equal?
\newblock \emph{Ethics}, 98\penalty0 (3):\penalty0 518--533, 1988.
\newblock ISSN 00141704, 1539297X.
\newblock URL \url{http://www.jstor.org/stable/2380965}.

\bibitem[Dwork et~al.(2012)Dwork, Hardt, Pitassi, Reingold, and
  Zemel]{dwork_awareness}
Cynthia Dwork, Moritz Hardt, Toniann Pitassi, Omer Reingold, and Richard~S.
  Zemel.
\newblock Fairness through awareness.
\newblock In Shafi Goldwasser, editor, \emph{Innovations in Theoretical
  Computer Science 2012, Cambridge, MA, USA, January 8-10, 2012}, pages
  214--226. {ACM}, 2012.
\newblock \doi{10.1145/2090236.2090255}.
\newblock URL \url{https://doi.org/10.1145/2090236.2090255}.

\bibitem[Agan and Starr(2018)]{agan2018ban}
Amanda Agan and Sonja Starr.
\newblock Ban the box, criminal records, and racial discrimination: A field
  experiment.
\newblock \emph{The Quarterly Journal of Economics}, 133\penalty0 (1):\penalty0
  191--235, 2018.

\bibitem[Lipton et~al.(2018)Lipton, McAuley, and Chouldechova]{Lipton_DLP}
Zachary Lipton, Julian McAuley, and Alexandra Chouldechova.
\newblock Does mitigating ml\textquotesingle s impact disparity require
  treatment disparity?
\newblock In S.~Bengio, H.~Wallach, H.~Larochelle, K.~Grauman, N.~Cesa-Bianchi,
  and R.~Garnett, editors, \emph{Advances in Neural Information Processing
  Systems}, volume~31. Curran Associates, Inc., 2018.
\newblock URL
  \url{https://proceedings.neurips.cc/paper/2018/file/8e0384779e58ce2af40eb365b318cc32-Paper.pdf}.

\bibitem[Hardt et~al.(2016)Hardt, Price, and Srebro]{hardt_EOP2016}
Moritz Hardt, Eric Price, and Nati Srebro.
\newblock Equality of opportunity in supervised learning.
\newblock In D.~Lee, M.~Sugiyama, U.~Luxburg, I.~Guyon, and R.~Garnett,
  editors, \emph{Advances in Neural Information Processing Systems}, volume~29,
  pages 3315--3323. Curran Associates, Inc., 2016.
\newblock URL
  \url{https://proceedings.neurips.cc/paper/2016/file/9d2682367c3935defcb1f9e247a97c0d-Paper.pdf}.

\bibitem[Chouldechova(2017)]{chouldechova_impossibility}
Alexandra Chouldechova.
\newblock Fair prediction with disparate impact: A study of bias in recidivism
  prediction instruments.
\newblock \emph{Big data}, 5\penalty0 (2):\penalty0 153--163, 2017.

\bibitem[Kleinberg et~al.(2016)Kleinberg, Mullainathan, and
  Raghavan]{Kleinberg_impossibility}
Jon~M. Kleinberg, Sendhil Mullainathan, and Manish Raghavan.
\newblock Inherent trade-offs in the fair determination of risk scores.
\newblock \emph{CoRR}, abs/1609.05807, 2016.
\newblock URL \url{http://arxiv.org/abs/1609.05807}.

\bibitem[Chouldechova and Roth(2020)]{chouldechova_frontiers}
Alexandra Chouldechova and Aaron Roth.
\newblock A snapshot of the frontiers of fairness in machine learning.
\newblock \emph{Commun. {ACM}}, 63\penalty0 (5):\penalty0 82--89, 2020.
\newblock \doi{10.1145/3376898}.
\newblock URL \url{https://doi.org/10.1145/3376898}.

\bibitem[Angwin et~al.(2016)Angwin, Larson, Mattu, Kirchner, and
  ProPublica]{compas_propublica}
Julia Angwin, Jeff Larson, Surya Mattu, Lauren Kirchner, and ProPublica.
\newblock Machine bias.
\newblock \emph{.}, 2016.

\bibitem[Department(2016)]{northpoine}
Northpointe Inc.~Research Department.
\newblock Compas risk scales: Demonstrating accuracy equity and predictive
  parity.
\newblock \emph{.}, 2016.

\bibitem[Heidari et~al.(2019)Heidari, Loi, Gummadi, and Krause]{heidari_EOP}
Hoda Heidari, Michele Loi, Krishna~P. Gummadi, and Andreas Krause.
\newblock A moral framework for understanding fair {ML} through economic models
  of equality of opportunity.
\newblock In \emph{Proceedings of the Conference on Fairness, Accountability,
  and Transparency, FAT* 2019, Atlanta, GA, USA, January 29-31, 2019}, pages
  181--190. {ACM}, 2019.
\newblock \doi{10.1145/3287560.3287584}.
\newblock URL \url{https://doi.org/10.1145/3287560.3287584}.

\bibitem[Liu et~al.(2021)Liu, Shafi, Fleisher, Eliassi-Rad, and
  Alfeld]{Liu2021Rawlsnet}
David Liu, Zohair Shafi, Will Fleisher, Tina Eliassi-Rad, and Scott Alfeld.
\newblock Rawlsnet: Altering bayesian networks to encode rawlsian fair equality
  of opportunity.
\newblock \emph{Proceedings of the 2021 AAAI/ACM Conference on AI, Ethics, and
  Society}, 2021.

\bibitem[Liu et~al.(2018)Liu, Dean, Rolf, Simchowitz, and
  Hardt]{liu2018delayed}
Lydia~T Liu, Sarah Dean, Esther Rolf, Max Simchowitz, and Moritz Hardt.
\newblock Delayed impact of fair machine learning.
\newblock In \emph{International Conference on Machine Learning}, pages
  3150--3158. PMLR, 2018.

\bibitem[Kannan et~al.(2019)Kannan, Roth, and Ziani]{downstream_kannan}
Sampath Kannan, Aaron Roth, and Juba Ziani.
\newblock Downstream effects of affirmative action.
\newblock In \emph{2019 ACM Conference on Fairness, Accountability, and
  Transparency}, FAT* '19, page 240–248, New York, NY, USA, 2019. Association
  for Computing Machinery.
\newblock ISBN 9781450361255.
\newblock \doi{10.1145/3287560.3287578}.
\newblock URL \url{https://doi.org/10.1145/3287560.3287578}.

\bibitem[Blum et~al.(2022)Blum, Stangl, and Vakilian]{multistage-Facct22}
Avrim Blum, Kevin Stangl, and Ali Vakilian.
\newblock Multi stage screening: Enforcing fairness and maximizing efficiency
  in a pre-existing pipeline.
\newblock In \emph{2022 ACM Conference on Fairness, Accountability, and
  Transparency}, FAccT '22, page 1178–1193, New York, NY, USA, 2022.
  Association for Computing Machinery.
\newblock ISBN 9781450393522.
\newblock \doi{10.1145/3531146.3533178}.
\newblock URL \url{https://doi.org/10.1145/3531146.3533178}.

\end{thebibliography}
